\documentstyle[twoside,fleqn,espcrc2,psfig]{article}

\newcommand{\AmS}{{\protect\the\textfont2
  A\kern-.1667em\lower.5ex\hbox{M}\kern-.125emS}}

\title{Monopoles and the Chiral Phase Transition in $SU(2)$ Lattice 
Gauge Theory}

\author{Roy Wensley \address{Department of Mathematical Sciences, Saint
Mary's College, Moraga, California, 94575, USA}
        \thanks{This work was supported in part by the National Science
Foundation, Grant No. PHY-9403869, the Higher Education Funding
Council for Wales, and the Saint Mary's College Faculty
Development Fund.}}
       
\begin{document}

\begin{abstract}
In the quenched approximation we use the abelian and monopole fields from
abelian projection in
SU(2) lattice gauge theory to  numerically compute the value of the chiral
condensate.  The condensate calculated using abelian projection is observed 
to vanish at the same critical temperature as the full SU(2) theory predicts.
\end{abstract}

\maketitle

\section{Introduction}

%
Calculations of the $SU(2)$ string tension using abelian projected
fields and their monopole contributions have given good evidence that 
confinement in non-abelian theories can be explained by abelian degrees of 
freedom and that the monopoles in these configurations are the confinement
mechanism\cite{banks77}-\cite{shiba94}.
Color confinement is the not the only non-perturbative effect to be 
understood in gauge theories.  Chiral symmetry breaking is well known to
have no perturbative explanation and it seems clear that presence of a
non-zero chiral condensate is related
to topology in non-abelian theories\cite{thooft76,hands90}.
Is it possible for the effective abelian theory, which seems to give an 
explanation of confinement in terms of magnetic monopoles, to explain the
chiral phase transition?  In this paper we present results which demonstrate 
that the chiral phase transition in quenched $SU(2)$ lattice gauge theory
is reproduced in abelian projected gauge fields after fixing to the maximal
abelian gauge.

\section{Calculation of the Chiral Condensate}

The order parameter used to study the chiral symmetry transition is
the chiral condensate defined by
$$\langle \bar{\psi}\psi(m) \rangle = {1 \over V} {\rm Tr} (D~\hspace{-10pt} / (U)+m)^{-1},$$
where $V$ is the lattice volume, $m$ is the input mass of the quenched fermion,
and $U$ is the lattice gauge field.  Spontaneous chiral symmetry breaking
is observed in the limit $V\rightarrow \infty$ 
when $\langle\bar{\psi}\psi(0)\rangle\not= 0$, i.~e. when the
chiral condensate remains finite at zero bare quark mass.  To study the chiral 
phase transition numerically, it is most straightforward to 
use staggered fermions, since chiral symmetry is maintained explicitly on 
the lattice.   Using staggered fermions leads to the Dirac eigenvalue
equation

\begin{eqnarray}
iD \hspace{-.25cm}/ [U] \psi & = & \sum_\mu{i\eta_\mu(x) \over 2}
( U_\mu(x)\psi(x+\hat{\mu}) \nonumber \\ 
&  & - U^\dagger_\mu(x-\hat{\mu})\psi(x-\hat{\mu})) 
\nonumber \\ & = & \lambda_n\psi(x) \label{stag}
\end{eqnarray}

The calculation of the chiral condensate is done in the following way:
First, we use the Lanczos method to calculate a small set of the 
lowest eigenvalues of 
Eqn.~(\ref{stag}).  These lowest eigenvalues are then used to compute
the spectral density function $\rho(\lambda)$.  Finally, the Banks-Casher
formula\cite{banks80}
$$\langle \bar{\psi}\psi(0)\rangle=\pi\rho(0)$$
is used to extract the chiral condensate at zero bare quark mass.  

In Eq.~(1) The link variable 
$U_\mu(x)$ in Eq.~(1) can come from any gauge group.  In the work done
here, Eq.~(1) is used to compute eigenvalues from both $SU(2)$ configurations 
and from abelian configurations generated using abelian projection. 

\section{The Monte Carlo Calculations}

$SU(2)$ gauge field configurations were generated using
a heatbath Monte Carlo simulation on a $16 \times N_\tau$ lattice
at $\beta=2.5115$ for values of $N_\tau=$ 4, 6, 8, 12, and 16. 
After allowing for 5000 equilibrium updates, 100 configurations separated by
40 updates were used for measurements.  The abelian projection was done by 
fixing to the maximum abelian gauge.  The method used for gauge fixing  and
extracting abelian fields is 
described in Ref.~\cite{stack94}. 

\subsection{Boundary Conditions}

In computing fermionic observables at finite temperature it is important to 
consider the boundary conditions.  To enforce Fermi statistics for the
fermionic fields on the lattice, anti-periodic boundary conditions in the
$\tau$-direction should be used.  For the fundamental gauge group 
representation it is known (numerically) that the chiral restoration and 
deconfinement transitions occur at the same critical temperature\cite{kogut83}.
Chiral restoration is signaled by a vanishing chiral condensate while
deconfinement is signaled by a non-vanishing Polyakov line value.  
At the value of $\beta$ used in this study, the
phase transitions are known to occur at $N_{\tau}=8$\cite{finberg}.

The deconfinement phase in pure $SU(2)$ is also accompanied by the spontaneous
breaking of a discrete global $Z_2$ symmetry.  This breaking of the
$Z_2$ symmetry is reflected in the sign of the average value of the Polyakov
line $\langle P\rangle$.  The global sign of the Polyakov line will affect the
boundary condition when solving Eq.~(1).  In order to maintain anti-periodic
boundary conditions (APBC) it is necessary to implement periodic boundary
conditions (PBC) when $\langle P \rangle < 0$ and APBC when 
$\langle P \rangle > 0$ (see Ref.~\cite{kogut83} for a discussion).
The subtlety with boundary conditions is expected to be an issue 
only for the quenched approximation and only in the chirally
symmetric phase, i.~e.~ only for $N_\tau<8$.

\section{Results}

The spectral density function was calculated for the full $SU(2)$, 
abelian projected, and monopole gauge fields.  The 25 lowest eigenvalues
from the 100 configurations at each $N_\tau$ were used.  The average
value of the Polykov line $\langle P \rangle$ was monitored for each
configuration.
In all cases it was found that the value ${\rm sgn}(\langle P \rangle)$ was
the same for the full $SU(2)$, abelian, and monopole fields of each
configuration.  For the case $N_\tau=4$ it was found that 
$\langle P \rangle < 0$, thus PBC were used in computing the eigenvalues.
While, for $N_\tau=6$ it was found that $\langle P \rangle > 0$, and so
APBC were used.

In Fig.~(1), the
results for $N_\tau=4$ are shown.  It seems clear that all three 
functions are approaching the same intercept, however the detailed behavior
does not agree.  Here we are considering the physics of the chiral
condensate and so we are only interested in the intercept.  In Fig.~(2), 
$\rho(\lambda)$ is shown for the abelian fields at each value of
$N_\tau$ used that corresponds to a finite temperature.   
A constant+linear+quadratic fit was used to characterize
the functions, and the intercept, $\rho(0)$, was extracted as a measure
of the condensate.  
\begin{figure}[hbt]
\psfig{file=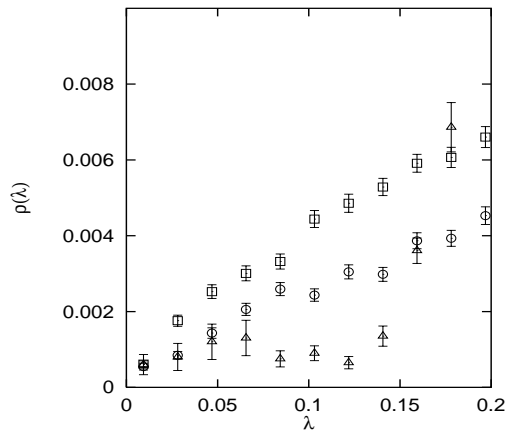,width=7cm,height=6.cm,angle=-90}
\caption{A comparison of the spectral density function at $N_\tau=4$  
for full $SU(2)$~(triangles), abelian~(squares), and 
monopole~(circles) fields.}  
\label{comp}
\end{figure}
\begin{figure}[hbt]
\psfig{file=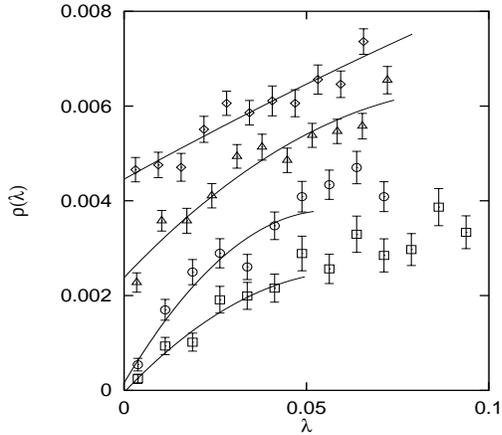,width=7cm,height=6.cm,angle=-90}
\caption{The spectral density function $\rho(\lambda)$ for small eigenvalues
with $N_\tau=$ 4 (squares), 6 (circles), 8 (triangles), and 12 (diamonds).  
The results for zero temperature ($N_\tau=16$) are not shown for the sake
of clarity, as they lie on top of the $N_{\tau}=12$ points.}
\label{fig1}
\end{figure}

As a check that the boundary conditions used did indeed make a difference
for configurations above the critical temperature, calculations at 
$N_\tau = 4$ using
APBC were also done.  In these calculations, the behavior of $\rho(\lambda)$ 
was found to
be more like the symmetrically broken case, and the intercept was
consistent with a non-zero value.  The value of
$\rho(0)$ for the APBC was found to be .0048(2).  This can be
compared to the value found for the PBC case (shown in Fig.~(1)) which was
found to be -.00004(20), which is consistent with zero.

In Fig.~(3), the value of $\rho(0)$ (using the correct boundary conditions)
as a function of $N_\tau$ is
presented.  It is clear from the figure that chiral symmetry is
restored at $N_\tau=8$.  This corresponds to the accepted critical
temperature of previous studies using full $SU(2)$ gauge fields\cite{finberg}.
\begin{figure}
\psfig{file=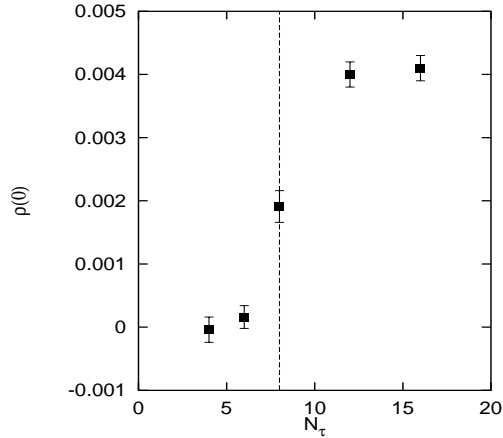,width=7cm,height=6.cm,angle=-90}
\caption{The value of $\rho(0)$ as a function of inverse temperature
$N_\tau$.  The dashed line is the accepted location of the phase transition.}
\label{fig2}
\end{figure}

\section{Conclusions}

The results presented indicate that the chiral phase transition can
be observed using the abelian projected fields in $SU(2)$.    This
gives evidence that non-perturbative effects in non-abelian theories
may be explained by an effective abelian theory, and thus ultimately 
tied to magnetic monopoles.

\end{document}